\documentclass{mn2e}

\newif\ifAMStwofonts

\usepackage{graphics}
\newcommand{\petunia}{Sk-69$\degr$194 }
\newcommand{\ms}{\, \mathrm M_{\sun}}
\newcommand{\rs}{\, \mathrm R_{\sun}}

\title{
The intriguing case of Sk-69$\degr$194
}
\author[Pablo G. Ostrov]
{Pablo G. Ostrov\thanks{E-mail: ostrov@fcaglp.edu.ar}
\\ Facultad de Ciencias Astron\'omicas y Geof\'{\i}sicas,
Paseo del Bosque S/N (1900) La Plata, Argentina \\
}
\date{}

\pagerange{\pageref{firstpage}--\pageref{lastpage}}

\begin{document}

\maketitle

\label{firstpage}

\begin{abstract}
In this paper we report the discovery of eclipses in the early-type
Magellanic binary \petunia. We derive an ephemeris for this system, and 
present a CCD $V$ light curve together with CCD spectroscopic observations. 
We also briefly discuss the nature of the binary components.
\end{abstract}


\section{Introduction}

\petunia ($\alpha=5:34:36$, $\delta=-69:45:37$, J2000)
is one of the brightest stars in the OB association LH\,81
 (Lucke \& Hodge 1970), in the Large Magellanic Cloud. Westerlund 
\cite{weste} studied that region, and obtained for \petunia (which he 
identified as the star number 10 in Table 28) $V=12.30$ and $B-V=-0.15$ 
(notice that Figs. 7 and 8 are interchanged in Westerlund's work, i.e. 
\petunia is the star number 10 in his Fig. 7). In its regard, he just pointed 
out that probably it was a variable star.
The name  \petunia comes from the Sanduleak's objective prism
survey (Sanduleak 1969). Later on, Isserstedt \cite{isser} performed 
$UBV$ photo-electric photometry, obtaining $V=11.98:$, $B-V=-0.08$ and 
$U-B=-0.96$. He also suspected that \petunia was a photometric variable.
More recently, Massey et al. \cite{mass} classified \petunia as B0\,I+WN.

LH\, 81 is one of the regions that we observed when searching new 
eclipsing binaries. In this paper we announce that \petunia is an eclipsing
binary system and present its light curve and radial velocity measurements. 
We have also tried to model our data by means of the Wilson-Devinney 
(hereafter W-D) programs (Wilson \& Devinney 1971, Wilson 1990, Wilson 1993).
Based on the spectroscopic data and the results of the W-D model, we discuss
the nature of the binary components.

\section{Observations and Reductions}

CCD $V$ photometry of \petunia was acquired during four observing runs between
1998 and 2001, with the 2.15-m telescope at CASLEO\footnote{
Complejo Astron\'omico El Leoncito, operated under agreement between 
the Consejo Nacional de Investigaciones Cient\'{\i}ficas y T\'ecnicas de la 
Rep\'ublica Argentina and the National Universities of La Plata, C\'ordoba and 
San Juan.}. A focal reducer was used, yielding a circular field of diameter 
about 9 arcmin and a scale of 0.813 arcsec px$^{-1}$.
The reduction of the data was performed by means of IRAF\footnote{
IRAF software is distributed by NOAO, operated by AURA for NSF.}
routines.

The CCD spectrograms were obtained with a REOSC 
\footnote{The REOSC spectrograph is on long term loan from the Li\`ege 
Observatory.}
spectrograph used in its single dispersion mode, attached to the same 
telescope between 2000 and 2001. A 600 grooves mm$^{-1}$ grating was used, 
giving a reciprocal dispersion of 1.63 \AA~px$^{-1}$. Our spectrograms
comprise the range from about 3900 to 5500 \AA.
The reductions and analysis of the 
spectroscopic observations were also performed by means of IRAF.

Both the photometric and the spectroscopic observations were acquired by means
of a CCD Tek $1024 \times 1024$ detector.

\section{Photometry}

Aperture photometry
was carried out using a stand-alone version of DAOPHOT (Stetson 1987, 1991).
In order to tie all the instrumental magnitudes to a common instrumental 
system, we used about 15 stars in the field of \petunia as local standards.
Just after the first observing run, the photometric data showed deep 
eclipses, of the order of a half magnitude. 

During a photometric night, in December 5, 1998, we also acquired $B$ and $R$
frames of the \petunia field, together with  photometric standards of the 
selected areas SA-92 and SA-98 (Landolt 1992), with the aim of linking the 
instrumental and standard photometric systems. 

From the observations
of that night, we derive the following magnitudes and colours for \petunia:
$$
V=12.378 \pm 0.010
$$
$$
B-V=0.013 \pm 0.016
$$
$$
V-R=-0.010 \pm 0.014
$$
at HJD=2451153.813. This corresponds to $\phi=0.534$, therefore the colour 
indexes must be considered with caution. Besides, \petunia
showed to have slight variations in the shape of its eclipses (see below).
The magnitude corresponding to $\phi=0.25$ is $V=11.83$.
Table \ref{foto} displays our $V$ aperture photometry (transformed to the
standard system), together with the 
internal errors derived from the local standards, seeing and airmass.

\begin{table*}  
\caption[]{$V$ light photometry of \petunia}
\label{foto}
  
\begin{tabular}{rrrrlrrrrl}  
\hline  
\noalign{\smallskip}
  
 HJD       &$    V    $&$ \sigma_{\rm i}$&   fwhm    &$  X    $&
 HJD       &$    V    $&$\sigma_{\rm i}$&   fwhm    &$  X    $\cr  
$  2450000+ $&           &               &$  \arcsec  $&               &
$  2450000+ $&           &               &$  \arcsec  $&               \cr
\noalign{\smallskip}  
\hline  
\noalign{\smallskip}  

$  1149.859 $&$  11.881 $&$   0.011 $&$   1.98 $&$   1.44 $&$  1504.621 $&$  11.883 $&$   0.007 $&$   2.76 $&$   1.45 $\cr
$  1151.605 $&$  11.944 $&$   0.007 $&$   1.96 $&$   1.41 $&$  1504.729 $&$  11.873 $&$   0.007 $&$   3.27 $&$   1.27 $\cr
$  1151.663 $&$  11.973 $&$   0.008 $&$   1.92 $&$   1.30 $&$  1504.768 $&$  11.868 $&$   0.007 $&$   2.68 $&$   1.27 $\cr
$  1151.706 $&$  11.968 $&$   0.008 $&$   1.94 $&$   1.27 $&$  1504.806 $&$  11.864 $&$   0.005 $&$   3.12 $&$   1.29 $\cr
$  1151.773 $&$  11.988 $&$   0.006 $&$   2.16 $&$   1.29 $&$  1504.834 $&$  11.861 $&$   0.007 $&$   2.96 $&$   1.33 $\cr
$  1151.844 $&$  12.005 $&$   0.005 $&$   2.09 $&$   1.41 $&$  1859.590 $&$  11.849 $&$   0.007 $&$   3.42 $&$   1.68 $\cr
$  1152.655 $&$  12.235 $&$   0.003 $&$   2.13 $&$   1.31 $&$  1859.607 $&$  11.843 $&$   0.010 $&$   3.50 $&$   1.60 $\cr
$  1152.701 $&$  12.249 $&$   0.009 $&$   1.83 $&$   1.27 $&$  1859.650 $&$  11.848 $&$   0.007 $&$   2.78 $&$   1.45 $\cr
$  1152.744 $&$  12.272 $&$   0.007 $&$   2.13 $&$   1.27 $&$  1859.685 $&$  11.849 $&$   0.009 $&$   3.14 $&$   1.36 $\cr
$  1152.793 $&$  12.288 $&$   0.008 $&$   2.19 $&$   1.32 $&$  1859.707 $&$  11.855 $&$   0.007 $&$   3.70 $&$   1.32 $\cr
$  1152.824 $&$  12.304 $&$   0.005 $&$   2.22 $&$   1.37 $&$  1859.740 $&$  11.852 $&$   0.028 $&$   5.70 $&$   1.29 $\cr
$  1152.860 $&$  12.313 $&$   0.009 $&$   2.06 $&$   1.46 $&$  1859.743 $&$  11.848 $&$   0.011 $&$   4.68 $&$   1.28 $\cr
$  1153.678 $&$  12.429 $&$   0.005 $&$   1.56 $&$   1.28 $&$  1859.778 $&$  11.852 $&$   0.008 $&$   5.36 $&$   1.27 $\cr
$  1153.729 $&$  12.411 $&$   0.005 $&$   1.69 $&$   1.27 $&$  1859.803 $&$  11.842 $&$   0.008 $&$   4.80 $&$   1.27 $\cr
$  1153.761 $&$  12.398 $&$   0.006 $&$   1.77 $&$   1.29 $&$  1859.841 $&$  11.838 $&$   0.016 $&$   4.53 $&$   1.30 $\cr
$  1153.811 $&$  12.378 $&$   0.006 $&$   2.40 $&$   1.35 $&$  1859.867 $&$  11.847 $&$   0.008 $&$   4.51 $&$   1.33 $\cr
$  1153.852 $&$  12.366 $&$   0.004 $&$   2.64 $&$   1.45 $&$  1860.610 $&$  11.907 $&$   0.009 $&$   4.99 $&$   1.58 $\cr
$  1154.699 $&$  12.111 $&$   0.007 $&$   1.99 $&$   1.27 $&$  1860.698 $&$  11.941 $&$   0.008 $&$   4.13 $&$   1.33 $\cr
$  1154.734 $&$  12.103 $&$   0.006 $&$   2.32 $&$   1.27 $&$  1860.771 $&$  11.945 $&$   0.008 $&$   3.77 $&$   1.27 $\cr
$  1154.773 $&$  12.095 $&$   0.006 $&$   2.93 $&$   1.30 $&$  1860.844 $&$  11.957 $&$   0.006 $&$   3.60 $&$   1.30 $\cr
$  1154.840 $&$  12.083 $&$   0.005 $&$   3.03 $&$   1.42 $&$  1861.612 $&$  12.199 $&$   0.009 $&$   3.77 $&$   1.56 $\cr
$  1155.648 $&$  11.909 $&$   0.007 $&$   2.47 $&$   1.31 $&$  1861.674 $&$  12.229 $&$   0.005 $&$   3.75 $&$   1.38 $\cr
$  1155.705 $&$  11.900 $&$   0.010 $&$   1.87 $&$   1.27 $&$  1861.803 $&$  12.285 $&$   0.003 $&$   3.47 $&$   1.27 $\cr
$  1155.741 $&$  11.899 $&$   0.008 $&$   1.81 $&$   1.28 $&$  1861.849 $&$  12.300 $&$   0.005 $&$   3.48 $&$   1.32 $\cr
$  1155.770 $&$  11.893 $&$   0.007 $&$   1.83 $&$   1.30 $&$  1862.583 $&$  12.363 $&$   0.021 $&$   4.06 $&$   1.67 $\cr
$  1155.820 $&$  11.883 $&$   0.005 $&$   2.09 $&$   1.38 $&$  1862.644 $&$  12.351 $&$   0.006 $&$   2.92 $&$   1.44 $\cr
$  1155.846 $&$  11.876 $&$   0.003 $&$   2.57 $&$   1.45 $&$  1862.711 $&$  12.322 $&$   0.008 $&$   3.37 $&$   1.31 $\cr
$  1499.849 $&$  11.964 $&$   0.005 $&$   2.88 $&$   1.33 $&$  1862.769 $&$  12.302 $&$   0.004 $&$   2.92 $&$   1.27 $\cr
$  1500.607 $&$  12.104 $&$   0.009 $&$   4.09 $&$   1.53 $&$  1862.847 $&$  12.284 $&$   0.005 $&$   3.58 $&$   1.31 $\cr
$  1500.671 $&$  12.128 $&$   0.005 $&$   3.44 $&$   1.36 $&$  1863.568 $&$  12.082 $&$   0.005 $&$   2.57 $&$   1.73 $\cr
$  1500.742 $&$  12.155 $&$   0.006 $&$   3.26 $&$   1.27 $&$  1863.744 $&$  12.045 $&$   0.007 $&$   2.94 $&$   1.28 $\cr
$  1500.812 $&$  12.174 $&$   0.006 $&$   4.07 $&$   1.29 $&$  1863.856 $&$  12.014 $&$   0.005 $&$   2.80 $&$   1.33 $\cr
$  1500.853 $&$  12.199 $&$   0.006 $&$   3.14 $&$   1.34 $&$  1864.605 $&$  11.890 $&$   0.007 $&$   1.85 $&$   1.55 $\cr
$  1501.559 $&$  12.405 $&$   0.007 $&$   2.82 $&$   1.74 $&$  1864.666 $&$  11.891 $&$   0.005 $&$   1.93 $&$   1.37 $\cr
$  1501.633 $&$  12.397 $&$   0.006 $&$   3.16 $&$   1.44 $&$  1864.713 $&$  11.874 $&$   0.006 $&$   2.13 $&$   1.30 $\cr
$  1501.686 $&$  12.395 $&$   0.004 $&$   3.26 $&$   1.33 $&$  1864.791 $&$  11.872 $&$   0.005 $&$   2.67 $&$   1.27 $\cr
$  1501.777 $&$  12.406 $&$   0.005 $&$   2.89 $&$   1.27 $&$  2212.743 $&$  11.986 $&$   0.007 $&$   2.96 $&$   1.32 $\cr
$  1501.821 $&$  12.411 $&$   0.006 $&$   2.78 $&$   1.30 $&$  2212.808 $&$  11.976 $&$   0.006 $&$   2.59 $&$   1.27 $\cr
$  1501.856 $&$  12.400 $&$   0.005 $&$   2.82 $&$   1.35 $&$  2212.857 $&$  11.964 $&$   0.006 $&$   2.89 $&$   1.28 $\cr
$  1502.658 $&$  12.138 $&$   0.007 $&$   2.78 $&$   1.37 $&$  2213.679 $&$  11.867 $&$   0.006 $&$   2.56 $&$   1.45 $\cr
$  1502.741 $&$  12.120 $&$   0.005 $&$   2.00 $&$   1.27 $&$  2213.740 $&$  11.861 $&$   0.007 $&$   2.18 $&$   1.32 $\cr
$  1502.814 $&$  12.110 $&$   0.005 $&$   2.14 $&$   1.29 $&$  2213.787 $&$  11.860 $&$   0.005 $&$   2.50 $&$   1.27 $\cr
$  1502.848 $&$  12.098 $&$   0.006 $&$   2.28 $&$   1.34 $&$  2213.843 $&$  11.848 $&$   0.008 $&$   2.49 $&$   1.27 $\cr
$  1503.672 $&$  11.973 $&$   0.005 $&$   2.56 $&$   1.34 $&$  2214.663 $&$  11.859 $&$   0.006 $&$   1.96 $&$   1.50 $\cr
$  1503.735 $&$  11.971 $&$   0.006 $&$   2.54 $&$   1.27 $&$  2214.724 $&$  11.870 $&$   0.019 $&$   2.92 $&$   1.34 $\cr
$  1503.789 $&$  11.959 $&$   0.005 $&$   2.94 $&$   1.28 $&$  2214.725 $&$  11.848 $&$   0.020 $&$   3.09 $&$   1.34 $\cr
$  1503.811 $&$  11.961 $&$   0.008 $&$   2.90 $&$   1.29 $&$  2214.772 $&$  11.863 $&$   0.025 $&$   2.57 $&$   1.28 $\cr
$  1503.827 $&$  11.949 $&$   0.004 $&$   3.14 $&$   1.31 $&$  2214.804 $&$  11.881 $&$   0.006 $&$   2.18 $&$   1.27 $\cr
$  1503.846 $&$  11.946 $&$   0.006 $&$   2.88 $&$   1.34 $&$  2214.840 $&$  11.893 $&$   0.006 $&$   2.07 $&$   1.27 $\cr
$  1504.565 $&$  11.878 $&$   0.005 $&$   3.63 $&$   1.67 $&$  2214.864 $&$  11.890 $&$   0.006 $&$   2.27 $&$   1.29 $\cr

\noalign{\smallskip}  
\hline  
\end{tabular}
\end{table*}  
%
%

\begin{figure}
  \resizebox{\hsize}{!}{\includegraphics{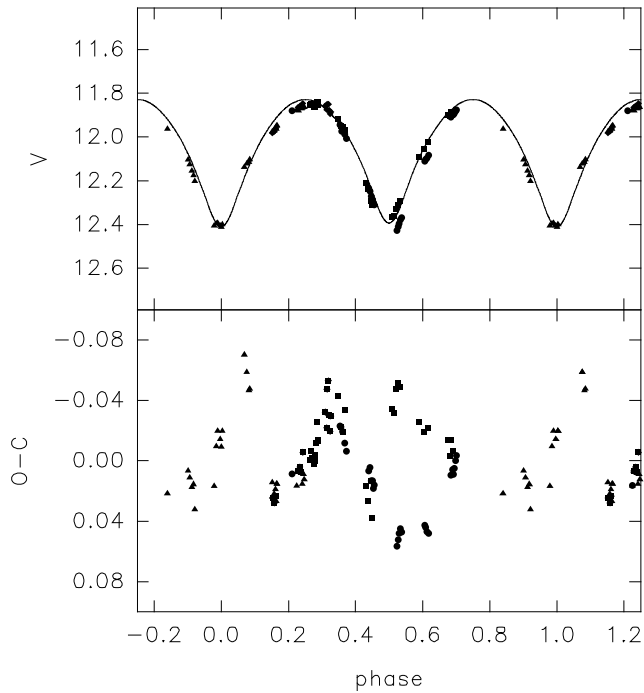}}
  \caption[]{The observed and modelled $V$ light curves for \petunia and their
concomitant O-C residuals. Circles, triangles, squares and diamonds stand for 
the data acquired during the observing runs in 1998, 1999, 2000 and 2001,
respectively.
}
        \label{lc}
\end{figure}

\section{Spectroscopy}

The spectrum of \petunia exhibits noticeable changes with phase. Only the 
spectrum of one of the stars is distinguishable.
Based on our data and ultraviolet observations downloaded from the IUE
\footnote{
Operated by the National Aeronautics and Space Administration, the European 
Space Agency and the Science and Engineering Research Council of the U.K.
}
archive, we classified the spectrum according to the classification criteria 
of Walborn \& Fitzpatrick \cite{walfitz}, corresponding  approximately to 
the type BN0\,Ia. That spectrum is easily measurable, and shows 
orbital velocity excursions of $\sim 425$ kms$^{-1}$ of amplitude. The 
emission in the region around N{\sc iii} 4634--He{\sc ii} 4686, very weak 
compared with which is usual in WN stars (as also pointed out Massey et al. 
2000), does not show any mensurable velocity variation as far as can be 
determined from our data. This spectral region also shows the 
more striking phase changes, due to the blend of the emission with the  
supergiant's absorption lines.

We determined the radial velocity at phase 0.016 (when the B supergiant is at
inferior conjunction) measuring several absorption lines. The velocities
corresponding to the other phases were derived by  cross correlation by means 
of the FXCOR IRAF task, using the spectra taken at phase 0.016 as template and
avoiding the range of 4500--4800 \AA.
The measured velocities are given in Table \ref{vr}.

\begin{figure*}
  \resizebox{12cm}{!}{\includegraphics{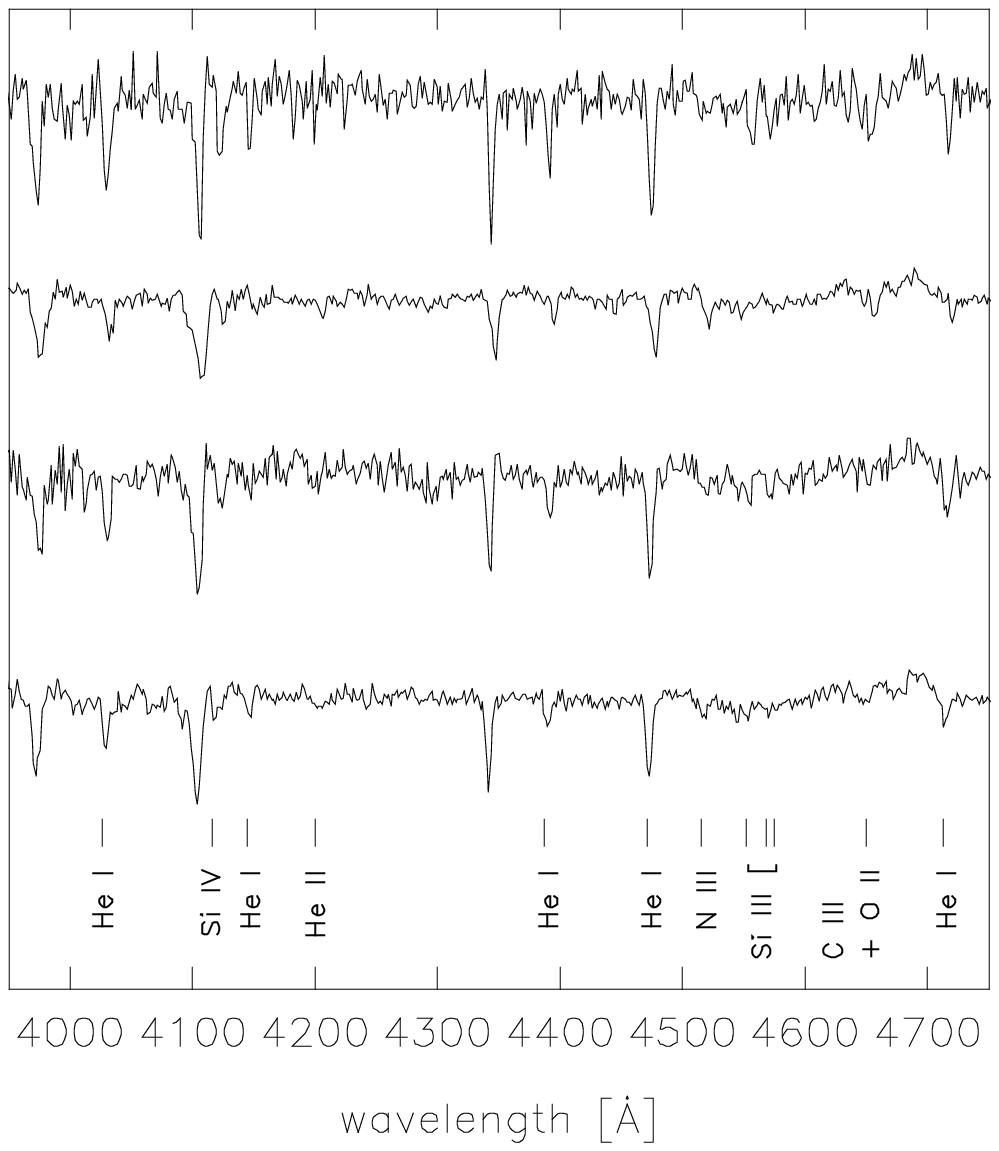}}
  \caption[]{
Normalised CCD spectra of \petunia obtained at CASLEO, corresponding (from 
top to bot) to phases 0.016, 0.284, 0.503 and 0.634, respectively.
}
 \label{spectra}
\end{figure*}

Four of the obtained spectra, corresponding to both conjunctions and both
quadratures, are displayed in Fig. \ref{spectra}. It is interesting to point
out that the N{\sc iii} 4515 absorption is more notorious in the spectra 
acquired near quadratures.

\begin{table}
\label{vr}
 \centering
  \caption{Radial velocities}
  \begin{tabular}{@{}lccc@{}}
\hline
HJD-2450000 & phase & B0\,I  \cr
& [km s$^{-1}$] & [km s$^{-1}$]  \cr
\hline
2253.711 & 0.504 & $170 \pm 21$  \cr
2286.613 & 0.195 & $482 \pm 30$  \cr
2287.704 & 0.284 & $429 \pm 23$  \cr
2293.633 & 0.769 & $82 \pm 26$  \cr
2296.655 & 0.016 & $246 \pm 16$  \cr
2328.651 & 0.634 & $106 \pm 20$  \cr
\hline
\end{tabular}
\end{table}

\section{Ephemeris}

It was not straightforward to derive an ephemeris only from the photometric 
observations. This was due to the fact that the shape of the light curve 
varied between 1998 and 2001, and to the intrinsic incompleteness of our 
data.  With the aid of our radial velocity measurements, we succeeded in 
deriving a period and a time of minimum,
$$
P=12.2252 \pm 0.0035
$$
$$T_0=2451501.815 \pm 0.012
$$
The phase-folded light curve is displayed in Fig. \ref{lc}. The
data obtained during different observing seasons are plotted using different
symbols, in order to show the shape variation. We emphasize that the data 
corresponding to secondary eclipse (observed during 1998 as well as during 
2000) was obtained, in both cases, during consecutive nights; i.e. the
discrepancy can not be ascribed to a period uncertainty.

\section{Light and Velocity Curve Modelling}

Being visible the spectrum of only one binary component, the derivation of
masses and absolute dimensions for this system is not straightforward.
However, and alternative approach suitable when the system present a 
Roche-lobe filling geometry and deep eclipses is to obtain the mass ratio 
solely from the light curve. We attempted to follow that approach, although 
the variation of 
the shape of the secondary eclipse between 1998 and 2000 increases the 
uncertainties of the solution.

Unfortunately, it was not possible to derive a photometric $q$ from our
light curve. We found solutions with $q$ ($q=M_2/M_1$, where $M_2$ corresponds
to the B supergiant)
ranging from 0.6 to 1.67, with
practically the same degree of significance. The solution with $q=1.67$
gives  a mass of $\sim 127 \ms$ for the B0\,I star, therefore can be 
disregarded. Only the solutions with lower mass ratio ($q \sim 0.7 \sim 0.8$)
provide masses and radii according whit what is expected for B supergiants.
However, for all cases, the solutions correspond to overcontact systems.
In these configurations, thermal contact between both components exists, hence 
the surface temperatures of the stars are similar. It is compatible with the 
light curve, that
shows eclipses of similar depth, but the very different spectra of the stars
suggest unlike physical conditions of their atmospheres. In spite of the
wide range of mass ratios of the possible solutions, the orbital inclination 
is $75 \pm 1 \degr$ for all of them. This inclination, together with the 
observed radial velocity curve, yields for $a_2$ a value of $49.1 \pm 4.3 \rs$.

\section{Discussion}

Massey et al. (2000) classified \petunia as B0\,I+WN, and pointed
out that the weakness of the WR emissions presumably is due to the continuum
being dominated by the B supergiant. The temperature inferred from the
light curve is compatible with that of a late WN, but all the solutions point 
to comparable magnitudes. In addition, no orbital movement is detected
in the emission, although it would be intrinsically difficult to measure due
to its weakness.

On weighing the observational data and the W-D modelling, we speculate about 
the alternative nature of the companion of the B supergiant:

\begin{itemize}

\item May be that it is truly a late WN star, with very weak emissions.
The fact that no orbital movement is detectable from He{\sc II} 4686 can be
ascribed to the difficulty of measuring the radial velocity of a broad 
emission superposed with the absorption lines of the other star.

\item It is possible that we are seeing a thick disk that hide one of the 
binary components. The He{\sc II} 4686 emission can be originated in the region
between (or surrounding) the stars.

\end{itemize}

Both hypothesis, though, do not agree with the overcontact configuration 
inferred from the W-D solutions. In one case, this contradiction can be 
ascribed to the interaction effects caused by the WR winds and radiation 
pressure (see, for example, Drechsel et al. 1995).
On the other hand, if we are seeing an accretion disk, the W-D model must
be considered with extreme caution, since the system's shape would be very 
different from the Roche-lobe geometry. Kondo et al. \cite{kondo} reviewed 
the main problems 
of close binary modelling, especially that concerning to Roche-lobe geometry.
\petunia is a system where such objections appear to be especially pertinent. 
Anyhow, it is interesting to mention that Walborn et
al. \cite{walborn} suggest three different ways that leads to the WN stage,
one of which being mass transfer binaries.
In fact, the
lack of discernible spectral features of the B companion can be explained if
we suppose that it is behind a thick stream of turbulent matter, hypothesis
that can also give account for the eclipse shape variation occurred between
1998 and 2001. 
A spectroscopic study encompassing a complete orbital period, at a higher 
resolution than that one can reach at CASLEO, could contribute to enlighten 
this issue.

The system remarkably resembles HD 163181 ($\equiv$V453 Scorpii)
(Josephs et al. 2001, and references there in), in which only 
one spectrum is visible (BN0.5 Ia). That system has a period of $\sim 12$ days 
with radial 
velocity excursions of nearly 400 km s$^-1$, also exhibiting enhanced
nitrogen abundance and signs of mass transfer.

From the derived values for $a_2$ and $i$, we have
$$
a=49.1(1+q) \rs
$$
$$
M_1=10.65(1+q)^2 \ms
$$
and
$$
M_2=10.65q(1+q)^2 \ms.
$$
Assuming for the B0 supergiant a mass of $\sim 25 \ms$ (Schmidt-Kaler 1982), 
the corresponding $q$ is $\sim 0.75$ and $M_1$ is $\sim 32.5 \ms$. However, it 
is important to have in mind that the
derived value for the inclination depends on the assumption of Roche-lobe
geometry, although its constant value for a wide range of possible mass ratios
suggests that it should not be very far from the actual value.

\begin{figure}
  \resizebox{\hsize}{!}{\includegraphics{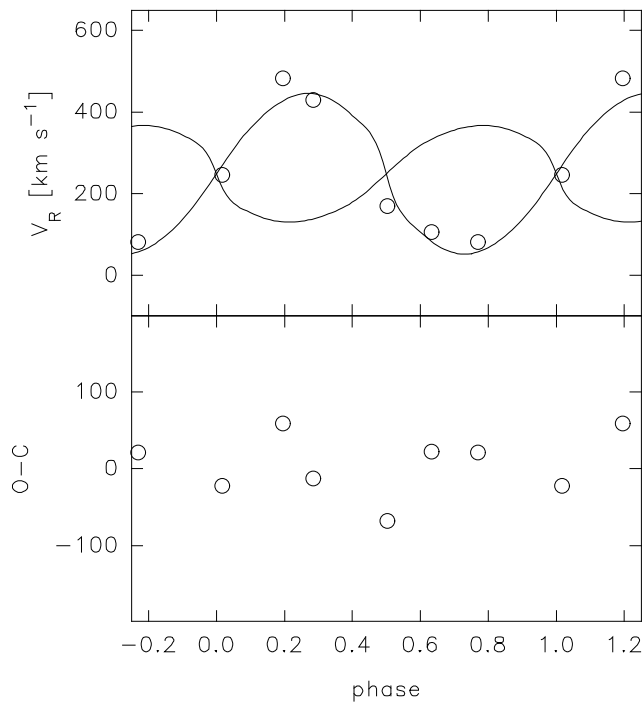}}
  \caption[]{Observed and modelled radial velocity curve for \petunia and
their O-C residuals.
}
        \label{vc}
\end{figure}


\section*{Acknowledgments}

I would like especially to thank Dr. Claudia Giordano for helping me 
with English.
I also wish to acknowledge Mariela Corti for the acquisition of two of the 
spectra, as an interchange of telescope time.
I also want to thank  Dr. Nidia Morrell and Dr Rodolfo Barb\'a for helpful 
comments, and Dr. Virpi Niemel\"a for her power of  motivation.  

The author acknowledges the use of the CCD and data acquisition system at 
CASLEO supported through a U.S. National Science Foundation grant AST-90-15827
to R. M. Rich. The focal reducer in use at CASLEO was kindly provided by Dr. M.
Shara.

\bsp

\label{lastpage}

\end{document}